# Synthesis and characterization of dense $Gd_2Ti_2O_7$ pyrochlore thin films deposited using RF magnetron sputtering

Cortney R. Kreller[1], Rangachary Mukundan[1], Eric L. Brosha[1], Todd Williamson[2], Terry Holesinger[3], James Valdez[4], Yong Q. Wang[4], and Blas P. Uberuaga[4]

[1]*MPA-11: Materials Synthesis and Integrated Devices, Los Alamos National Laboratory, Los Alamos, NM 87545*
[2]*C-NR: Nuclear and Radiochemistry, Los Alamos National Laboratory, Los Alamos, NM 87545*
[3]*MST-16: Nuclear Materials Science, Los Alamos National Laboratory, Los Alamos, NM 87545*
[4]*MST-8: Materials Science in Radiation and Dynamics Extremes, Los Alamos National Laboratory, Los Alamos, NM 87545*

Thin films of phase pure pyrochlore $Gd_2Ti_2O_7$ have been synthesized by RF magnetron sputtering. The films were prepared from oxide targets in 50%$O_2$/Ar atmosphere and deposited on 111 yttria-stabilized zirconia (YSZ) substrates at a temperature of 800°C. The pyrochlore structure was confirmed via grazing angle x-ray diffraction and selected area electron diffraction (SAED). Transmission electron microscopy (TEM) analysis also showed that the films were dense and of uniform thickness with surface roughness of approximately 8nm. The total conductivity measured with AC impedance spectroscopy was found to be independent of thickness and comparable in magnitude to that of bulk $Gd_2Ti_2O_7$. Differences were observed in the Arrhenius behavior between the bulk and thin film samples and are attributed to varying levels of background impurities.

**Introduction**
The solid solution $Gd_2Zr_xTi_{1-x}O_7$ has been studied extensively for its properties of fast ion conduction and radiation tolerance[1-3]. For both applications, a key enabling property is the ability for the cations to disorder. The ordered pyrochlore end member $Gd_2Ti_2O_7$ (GTO) is generally regarded as an insulating extrinsic ionic conductor, while the more easily disordered $Gd_2Zr_2O_7$ (GZO) exhibits intrinsic fast ion conduction.

The pyrochlore structure $A_2^{3+}B_2^{4+}O_7$ can be regarded as a defect-fluorite type structure with ordering on both the cation and anion sublattice yielding a superstructure with twice the lattice constant of the parent fluorite. The A and B site cations order in the <110> directions with each cation occupying alternate rows with the larger $A^{3+}$ cation 8-fold coordinated and the smaller $B^{4+}$ cation 6-fold coordinated. The structure consists of three distinct oxygen sites: 48f with two A and two B nearest neighbors, 8a with four A nearest neighbors and a vacant 8b site with four B nearest neighbors[4]. Previous studies have shown that, generally, the pyrochlore structure forms when $1.46<R_A/R_B<1.80$. Below 1.46 the cation radii are of similar enough size that anti-site order is introduced giving rise to the defect fluorite phase[5]. Disorder on the cation sublattice is accompanied, or perhaps preceded by[6], disorder on the anion sublattice. Disorder on the anion sublattice serves to increase the number of mobile oxygen vacancies through the formation of Frenkel defects consisting of a vacancy on the 48f site and an oxygen interstitial ion on the 8b site[7].

Numerous works have correlated increasing disorder with increasing ionic conductivity. For example, Moon and Tuller have shown that the isovalent substitution of Zr for Ti in $Gd_2Zr_xTi_{1-x}O_7$ results in an approximately 4 order of magnitude increase in conductivity at 600°C between x=0 and x=1.0[8]. Recent modeling results have indicated that cation disorder enhances ionic conductivity by increasing both the concentration of mobile charge carriers and their individual mobility[9]. Conversely, other researchers have shown that disorder serves to increase the amount mobile oxygen in the lattice, i.e. carrier concentration, but that order actually provides a preferential pathway for transport and thus a reduced activation energy. This is supported by the work of Diaz-Guillen et al. who examined the effect of homovalent lanthanide (Ln) A-site substitution on the pyrochlore phase of $Gd_{2-y}Ln_yZr_2O_7$ and demonstrated that activation energy decreased and conductivity increased, albeit slightly, with larger cations on the A-site (i.e. increasing order)[10].

Quantifying the extent of disorder in pyrochlore materials can be challenging. XRD patterns consist of two subsets of patterns. The first set arises from the defect fluorite structure, which is present in all materials regardless of order. The second subset is attributable to the pyrochlore superstructure. A common method for defining the extent of disorder is to use the ratio of the intensities of pyrochlore to fluorite peaks to calculate the cation order parameter[8, 11]. The superstructure peaks, however, are much weaker than those of the fluorite phase even in a fully ordered material, and decrease rapidly with increasing disorder[6]. Thus, the unambiguous determination of order is difficult from XRD data alone. Another factor complicating structural analysis in polycrystalline samples is the possibility of locally ordered phases in a matrix exhibiting long-range disorder[12, 13], or vice versa. An early study by van Dijk et al. used selected area electron diffraction (SAED) to identify ordered pyrochlore microdomains in a disordered fluorite structure[13]. A recent study by Blanchard et al. coupled neutron and X-ray powder diffraction with X-ray absorption near edge structure analysis of the Zr L-edge across the lanthanoid series of $Ln_2Zr_2O_7$ [14]. This study showed increasing local disorder with decreasing $R_A/R_B$ across the compositions identified as pyrochlore via neutron and x-ray diffraction, demonstrating the gradual nature of the transition from ordered pyrochlore to defect fluorite phase.

Unlike bulk polycrystalline materials, thin films provide a platform for precise phase control through epitaxial growth on substrates with appropriate lattice match. While the use of thin films to explore and exploit properties of fast ion conductors has gained a great deal of interest in recent literature[15, 16], this platform has not yet been utilized to control the structure and study the conductivity of pyrochlore materials. In this work we report conductivity measurements on thin films of the fully ordered pyrochlore $Gd_2Ti_2O_7$. To the best of our knowledge, this is the first synthesis of a dense thin film of a lanthanide-bearing pyrochlore.

**Experimental Details**
*RF magnetron sputtering-* Thin films were prepared using RF magnetron sputtering on 111 oriented single-crystal YSZ substrates (8 mol% $Y_2O_3$-stabilized $ZrO_2$, one-side polished, MTI corporation). The substrates were ultrasonically cleaned in DI water and Acetone and dried in air at 150°C for several hours. The YSZ substrates were affixed to the Ni faceplate of a heater box using silver epoxy (AREMCO). A stoichiometric $Gd_2Ti_2O_7$ target was synthesized from high purity $Gd_2O_3$ and $TiO_2$ powders. Powders were mixed in an inert atmosphere glove box, milled for 6 hours in a Retsch rotary mill, dried on a shlink line, and then calcined as a loose powder for 10 hours under $25\%O_2$/ Ar gas flow at 700°C. Powder was loaded into a 2.125 inch die and pressed at 15 lbs. This target was then sintered at 1575°C for 100 hours in an oxygen atmosphere furnace on a bed of loose powder in a high purity alumina boat. Shrinkage during the sintering cycle resulted in an approximately 2 inch diameter target. Before use, the target was sanded to remove any adhered powder from the sintering process. Phase purity of the $Gd_2Ti_2O_7$ target was confirmed by XRD. The sintered target was mounted in a 2-in. diameter copper RF magnetron sputter cup using Ag epoxy (Ted Paella). The heater box with the affixed substrate was mounted off-axis 3.5 inches from the target. The substrate was maintained at 800°C using a photo lamp heater bulb (Wiko). The films were grown in $50\%O_2$/balance Ar atmosphere with total pressure maintained at 25mTorr and at a power setting of 100W. These sputtering conditions yielded a deposition rate of roughly 17Å/min. Films with thicknesses of 480nm and 850nm were produced by 5 hour and 8-hour deposition times, respectively.

*Film Characterization-* Films were characterized by grazing incidence X-ray diffraction (GIXRD) and transmission electron microscopy (TEM). GIXRD measurements were performed using a Bruker AXS D8 Advance X-ray diffractometer, with a Cu-K$_\alpha$ radiation ($\lambda$=1.5406Å) X-ray source operating in $\theta$-$2\theta$ geometry, and at the fixed angle of incidence ($\gamma$) of 1° relative to the specimen surface. The X-ray diffractometer exit source was equipped with a Göebel mirror diffraction optic used to achieve a parallel beam condition that is highly sensitive to near surface features in thin films. The $\theta$-$2\theta$ scans were performed using a step size of 0.02° and a dwell time of 4 seconds per step. Diffracted intensities from the specimens were collected using a solid-state detector to suppress fluorescence from the rare-earth gadolinium constituent.
The structure of the films was examined in a FEI Titan 300kV image corrected (scanning) transmission electron microscope (S/TEM). Specimens for S/TEM analysis were prepared via a focused ion beam (FIB) process. The specimen was cut from the film, plucked and attached to a copper FIB grid, and then thinned to electron transparency.

*Electrochemical measurements*- Platinum ink (Heraeus 6082) was painted onto the unpolished side of the YSZ and used to adhere the substrate to a Pt foil counter electrode.  The ink was cured at 950°C for 30 minutes. A shadow mask (Photo Etch) was used to sputter platinum electrodes ranging from 0.1 to 1mm in diameter onto the GTO film.  A microprobe station (Cascade Microtech) equipped with a high temperature furnace stage (Linkham) was used to collect the impedance spectra.  Gold coated tungsten probe tips were contacted to the Pt foil counter electrode and a Pt working electrode on the surface of the GTO film.  AC impedance measurements were performed using a PARSTAT 2273 potentiostat/FRA in the frequency range of $10^6$-0.1 Hz as a function of temperature (500-900°C) in ambient gas environment.

*Bulk powder synthesis and characterization*- In order to compare the epitaxial thin films to bulk measurements, powders of GTO were synthesized from the oxide precursors $Gd_2O_3$ (Alfa Aesar, 99.999% purity) and $TiO_2$ (Alfa Aesar, 99.995% purity), by conventional ceramic solid state processing.  The oxide powders were first calcined at 1000°C for 24 hours, then ball-milled in a Spex Certiprep 8800 high energy ball-mill for 12 hours in $ZrO_2$ vials with $ZrO_2$ balls using a 10:1 ball to powder ratio for optimum milling and blending.  The resulting powders were then pressed at a pressure of 440 MPa in a 13 mm diameter stainless steel die. The resulting pellets were then sintered at 1200°C for 24 hours, then re-milled (same routine previously discussed), re-pressed and re-sintered at 1600°C for 24 hours. The measured (geometric) density of the final $Gd_2Ti_2O_7$ pellets were 80% of the theoretical values. Sections of ~1mm thickness were then polished to a mirror finish using diamond impregnated lapping films and oil as lubricant followed by a final polishing step using SYTON® HT-50 on a flocked twill cloth. X-ray diffraction measurements obtained from the samples showed them to be phase pure with the pyrochlore structure belonging to space group Fd-3m.  Despite best efforts to minimize contamination, the final cation stoichiometry was 0.04:0.96 Zr:Ti as identified by XRF (Thermo Scientific Quant'x) using a fundamental parameters model.  Al was also identified as a minor impurity via XRF. Platinum ink (Heraeus 6082) electrodes were painted on opposite sides of the pellet and cured at 950°C according to the manufacturers firing schedule. The pellets were compressed between two sheets of Pt foil using a standard spring-compression set-up.  The assembly was placed inside a quartz tube and electrochemical measurements were made as a function of temperature (600-1000°C) under ambient conditions in a tube furnace.

**Results and Discussion**
X-ray diffraction patterns of the bulk polycrystalline sample and the 480nm thin film are shown in **Figure 1** along with the JCP pattern for Gd2Ti2O7 pyrochlore (Fd-3m). The bulk polycrystalline sample exhibits the characteristic pyrochlore (111) and (331) superstructure peaks at ~15 and 37° 2θ, respectively. The signal to noise ratio is much lower for the thin film sample, however, the same superstructure peaks are observable in the thin film diffraction pattern, along with the family of peaks arising from the parent fluorite structure.

A TEM micrograph of a film deposited under the same sputtering conditions but for only 1 hour is shown in **Figure 2**a. The film is dense with no interior or interfacial porosity and exhibits a columnar structure. The epitaxial nature of the film growth from the YSZ interface can be seen in the high resolution image of **Figure 2**b. The dark/light regions in the substrate and film are indicative of residual strain within in the structure. The selected area diffraction pattern in **Figure 2**c clearly identifies the pyrochlore structure of the film. STEM line scans across the film show a uniform composition from the surface to the substrate interface. The average semi-quantitative film composition was 16.7:18.5:65.7 (Gd:Ti:O), consistent with the intended stoichiometry for a pyrochlore film.

A representative impedance spectrum obtained from a 2-probe through-plane measurement of the 480nm film using a 1mm diameter Pt-working electrode is shown in **Figure 3**. The low frequency arc is due to interfacial processes (Pt||GTO, GTO||YSZ, YSZ||Pt); the analysis of these processes was beyond the scope of this work. The intermediate frequency arc was of primary interest and is attributable to the GTO thin film. The high frequency intercept yields the ohmic resistance of the YSZ substrate, as electronic resistances were minimal. Fits to the high and intermediate frequency features were performed using ZView with an equivalent circuit of the form shown in **Figure 4**.

The conductivity of the YSZ substrate calculated from the fits of $R_{YSZ}$ was found to be in close agreement with literature values for all electrode diameters and both film thicknesses. The capacitance values obtained from fitting $C_{GTO}$ to the intermediate frequency arc scaled as expected with electrode area and film thickness yielding an average dielectric constant of 66, obtained from the relationship:

$$\varepsilon' = \frac{CL}{\varepsilon_o A} \qquad (1)$$

Where $L$ is the thickness of the sample, $A$ is the area of the smaller, working electrode and $\varepsilon_o$ is the permittivity of vacuum ($8.85 \times 10^{-12}$ Fm$^{-1}$). This value is in agreement with prior literature reports[17]. $R_{GTO}$ was then used to calculate the conductivity according to:

$$\sigma_{GTO} = \frac{1}{R_{GTO}} \frac{L}{A} \qquad (2)$$

The conductivity of the thin films along with measurements of bulk polycrystalline samples is shown in **Figure 5**. No dependence on film thickness is observed in the conductivity of the 480 and 850nm films indicating that any substrate-induced strain is not influencing the conductivity. This is expected since the strain is expected to relax within the first 100nm away from the interface. The magnitude of the conductivity is similar between the thin films and the bulk sample, however the Arrhenius response differs between the two.

The standard Arrhenius equation describing ionic conductivity is expressed as:

$$\sigma = \frac{\sigma_o}{T} Exp\left(\frac{-E_A}{kT}\right) \qquad (3)$$

where the pre-exponential constant $\sigma_o$ is proportional to the carrier concentration, $E_A$ is the activation energy of the bulk conduction process, and k is the Boltzmann constant.

The activation energy of 2.24eV found for the bulk polycrystalline GTO samples in this study is comparable to activation energies of 2.26eV and 2.09eV extracted from Kramer[18] and Moon[17], respectively, for the nominally pure GTO material. The pre-exponential constants are also consistent with these studies ($10^{7.4-8}$ SKcm$^{-1}$). The thin films exhibit lower activation energies of 1.58 and 1.41eV for the 480 and 850nm films respectively, as well as a lower pre-exponential constant ($10^{3.7-4.5}$ SKcm$^{-1}$), indicating both a reduced barrier to transport, as well as a lower concentration of mobile charge carriers. This explains why the conductivity of the thin films is higher than that of the bulk at lower temperatures, when the lower activation energy enables bulk diffusion, versus higher temperatures, where the higher concentration of mobile charge carriers leads to faster conduction in the bulk sample.

The differences in conductivity observed between the thin films and bulk polycrystalline material reported herein as well as in the literature is likely due to differences in extents of minor impurities in the material. The bulk material in this work was not pure GTO, but rather $Gd_2Ti_{0.96}Zr_{0.04}O_7$, and also contained Al as a minor impurity as identified via XRF. Other minor impurities may have been present as well but at ppm levels below the detection limit of XRF. The GTO material reported by Moon was analyzed using spark source mass spectrographic (SSMS) analysis and was found to contain a number of impurities including Ca, Al, Si, and Hf, as well as 90ppm of Zr[17]. Such impurity loadings could be regarded as inconsequential for an intrinsic ion conductor, however the charge compensation mechanisms of the material due to unintentional aliovalent dopants (such as $Ca^{2+}$ on the A-site or $Al^{3+}$ on the B-site) could have a significant effect on a poor conductor such as GTO. Recent molecular dynamics simulations have shown that introducing an intrinsic defect, either an oxygen vacancy or interstitial, into an ordered material results in a marked decrease in activation energy, whereas the effect is minimal in a disordered material where structural vacancies are already mobile[9]. Kramer reported on an additional GTO sample with different cation stoichiometry and higher background impurities[18] than the sample referenced above. This sample exhibited an activation energy of 0.94eV, which is lower than that of both the other bulk samples and the thin films, and a pre-exponential constant of $10^{3.14}$ SKcm$^{-1}$, which is comparable, but slightly less than that of the thin films. This would suggest that the thin films contain an impurity level with concentrations between the bulk sample reported herein and the material reported by Kramer, though further work is needed to quantify the impurities in the thin films.

As $Gd_2Ti_2O_7$, even with 5 mol% Zr, is well within the pyrochlore stability field, it is unlikely that any large microdomains of disorder within the material are contributing to the differences in the observed conductivity. However, it is possible that in addition to carrier creation via charge compensation, aliovalent impurity dopants may also serve to induce disorder, such as anti-site formation, locally within the lattice. Unfortunately it is not possible to de-convolute these two effects in GTO. Thin films of the intrinsic fast ion conductor $Gd_2Zr_xTi_{1-x}O_7$ with x>0.4 will be studied in the future in order to more fully explore the effects of order/disorder on oxygen transport in pyrochlore/defect fluorite materials.

**Conclusions**
Dense thin films of the ordered pyrochlore $Gd_2Ti_2O_7$ were, for the first time, synthesized via RF magnetron sputtering. GIXRD patterns of the films exhibited the characteristic \ pyrochlore superstructure diffraction peaks at ~15° and 38° 2θ, and SAED patterns showed sharp spots at positions corresponding to the pyrochlore structure. AC impedance measurements were used to isolate the impedance response of the GTO thin film, with the capacitance values scaling as expected with electrode area and film thickness. The total conductivity of the thin films was found to be independent of thickness and comparable in magnitude to that of the bulk polycrystalline material. Differences in the Arrhenius responses are likely attributable to varying background aliovalent impurities which have a marked influence on the carrier concentration and activation energy of migration in the extrinsic conductor GTO. The results of this study show that GTO thin films exhibit similar properties to the bulk polycrystalline material. The phase control enabled by epitaxial growth provides a means of tailoring the order in pyrochlore/defect-fluorite thin films. Future work will focus on thin films of the extended series of $Gd_2Zr_xTi_{1-x}O_7$ in order to better understand the relationship between order and transport.

**Acknowledgements**
This work was supported by the U.S. Department of Energy, Office of Science, Basic Energy Sciences, Materials Sciences and Engineering Division. LANL, an affirmative action/equal opportunity employer, is operated by Los Alamos National Security, LLC, for the National Nuclear Security Administration of the U.S. Department of Energy under contract DE-AC52-06NA25396.

**References**
1. Zhang, J.; Lian, J.; Zhang, F.; Wang, J.; Fuentes, A. F.; Ewing, R. C., Intrinsic Structural Disorder and Radiation Response of Nanocrystalline Gd2(Ti0.65Zr0.35)2O7 Pyrochlore. *Journal of Physical Chemistry C* **2010,** *114*, 11810-11815.
2. Zhang, J. M.; Lian, J.; Fuentes, A. F.; Zhang, F. X.; Lang, M.; Lu, F. Y.; Ewing, R. C., Enhanced radiation resistance of nanocrystalline pyrochlore Gd-2(Ti0.65Zr0.35)(2)O-7. *Applied Physics Letters* **2009,** *94* (24).

3.	Begg, B. D.; Hess, N. J.; McCready, D. E.; Thevuthasan, S.; Weber, W. J., Heavy-ion irradiation effects in Gd-2(Ti2-xZrx)O-7 pyrochlores. *Journal of Nuclear Materials* **2001,** *289* (1-2), 188-193.
4.	Burggraaf, A. J.; van Dijk, T.; Verkerk, M. J., Structure and Conductivity of Pyrochlore and Fluorite Type Solid Solutions. *Solid State Ionics* **1981,** *5*, 519-522.
5.	Subramanian, M. A.; Aravamundan, G.; Subba Rao, G. V., Oxide Pyrochlores- A Review. *Progress in Solid State Chemistry* **1983,** *15*, 55-143.
6.	Wuensch, B. J.; Eberman, K. W.; Heremans, C.; Ku, E. M.; Onnerud, P.; Yeo, E. M. E.; Haile, S. M.; Stalick, J. K.; Jorgensen, J. D., Connection between oxygen-ion conductivity of pyrochlore fuel-cell materials and structural change with composition and temperature. *Solid State Ionics* **2000,** *129*, 111-133.
7.	van Dijk, M. P.; Burggraaf, A. J.; Cormack, A. N.; Catlow, C. R. A., Defect Structures and Migration Mechanisms in Oxide Pyrochlores. *Solid State Ionics* **1985,** *17*, 159-167.
8.	Moon, P. K.; Tuller, H. L., Intrinsic Fast Oxygen Ionic Conductivity in the Gd2(ZrxTi1-x)2O7 and Y2(ZrxTi1-x)2O7 Pyrochlore Systems. *Materials Research Society Symposium Proceedings* **1989,** *135*, 149-163.
9.	Perriot, R.; Uberuaga, B. P., Structural vs. intrinsic carriers: contrasting effects of cation chemistry and disorder on ionic conductivity in pyrochlores. *Journal of Materials Chemistry A* **2015,** *3*, 11554-11565.
10.	Díaz-Guillén, J. A.; Fuentes, A. F.; Díaz-Guillén, M. R.; Almanza, J. M.; Santamaría, J.; León, C., The effect of homovalent A-site substitutions on the ionic conductivity of pyrochlore-type Gd2Zr2O7. *Journal of Power Sources* **2009,** *186* (2), 349-352.
11.	Zhang, F. X.; Lang, M.; Ewing, R. C., Atomic disorder in Gd2Zr2O7 pyrochlore. *Applied Physics Letters* **2015,** *106* (19).
12.	Whittle, K. R.; Cranswick, L. M. D.; Redfern, S. A. T.; Swainson, I. P.; Lumpkin, G. R., Lanthanum pyrochlores and the effect of yttrium addition in the systems La2-xYxZr2O7 and La2-xYxHf2O7. *Journal of Solid State Chemistry* **2009,** *182* (3), 442-450.
13.	Vandijk, M. P.; Mijlhoff, F. C.; Burggraaf, A. J., PYROCHLORE MICRODOMAIN FORMATION IN FLUORITE OXIDES. *Journal of Solid State Chemistry* **1986,** *62* (3), 377-385.
14.	Blanchard, P. E.; Clements, R.; Kennedy, B. J.; Ling, C. D.; Reynolds, E.; Avdeev, M.; Stampfl, A. P.; Zhang, Z.; Jang, L. Y., Does local disorder occur in the pyrochlore zirconates? *Inorg Chem* **2012,** *51* (24), 13237-44.
15.	Rupp, J. L. M., Ionic diffusion as a matter of lattice-strain for electroceramic thin films. *Solid State Ionics* **2012,** *207*, 1-13.
16.	García-Barriocanal, J.; Rivera-Calzada, A.; Varela, M.; Sefrioui, Z.; Iborra, E.; Leaon, C.; Pennycook, S. J.; Santamaria, J., Colossal Ionic Conductivity at Interfaces of Epitaxial ZrO2:Y2O3/SrTiO3 Heterostructures. *Science* **2008,** *231*, 676-680.
17.	Moon, P. K. Electrical conductivity and structural disorder in Gd2Ti2O7-Gd2Zr2O7 and Y2Ti2O7-Y2Zr2O7 solid solutions. Massachusetts Institute of Technology, Cambridge, 1988.

18.     Kramer, S. A. Mixed ionic-electronic conduction in rare earth titanate/zirconate pyrochlore compounds. Massachusetts Institute of Technology, 1994.

**Figures**

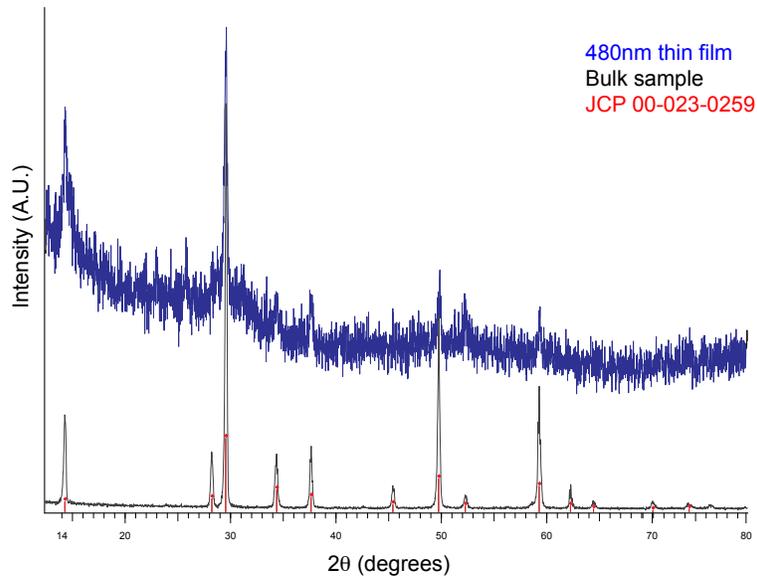

Figure 1:GIXRD pattern of 480nm thin film (blue), Bulk polycrystalline material (black) and JCP pattern for pyrochlore (Fd-3m) $Gd_2Ti_2O_7$ (red).

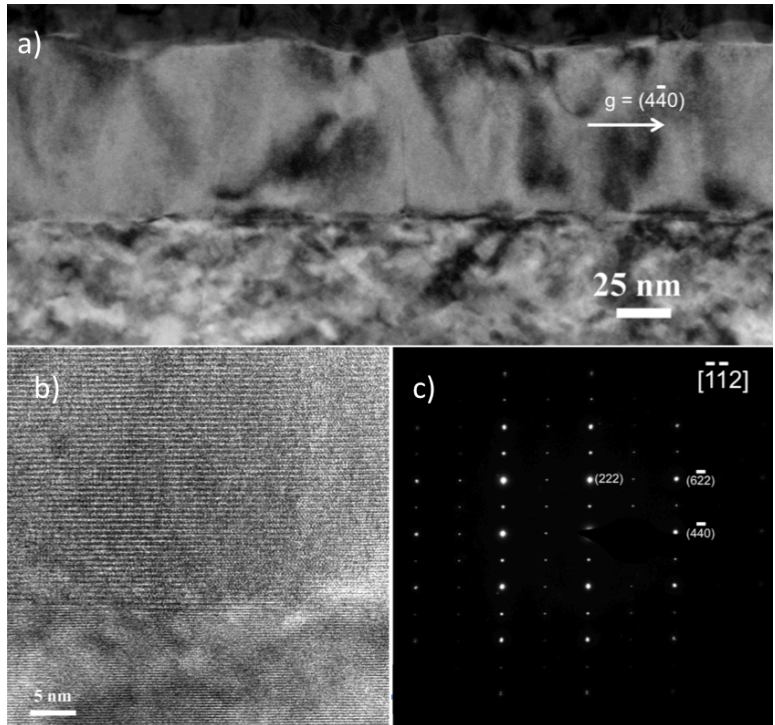

Figure 2: a) TEM of GTO thin film on 111 oriented YSZ substrate b) High resolution micrograph showing epitaxy of GTO film with YSZ substrate and c) SAED pattern showing sharp spots corresponding to pyrochlore structure.

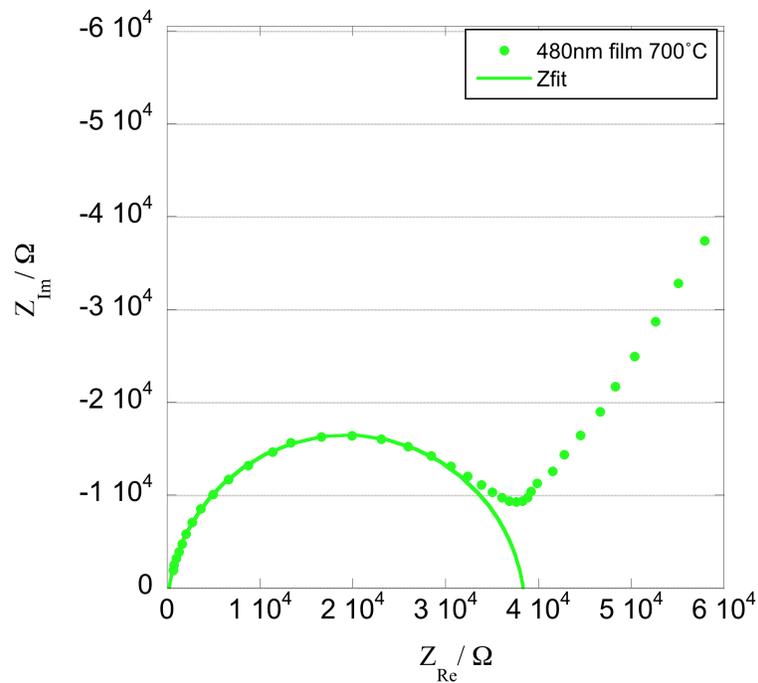

Figure 3: Impedance spectra obtained for 480nm film at 700°C using 2-probe through plane measurements between a 1mm diameter working electrode and the counter electrode (data points) and the corresponding fit obtained from Zview using the equivalent circuit shown in Figure 4.

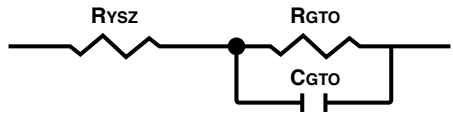

Figure 4: Equivalent circuit used to fit the high and intermediate frequency features of the impedance spectra of GTO thin films on YSZ substrates.

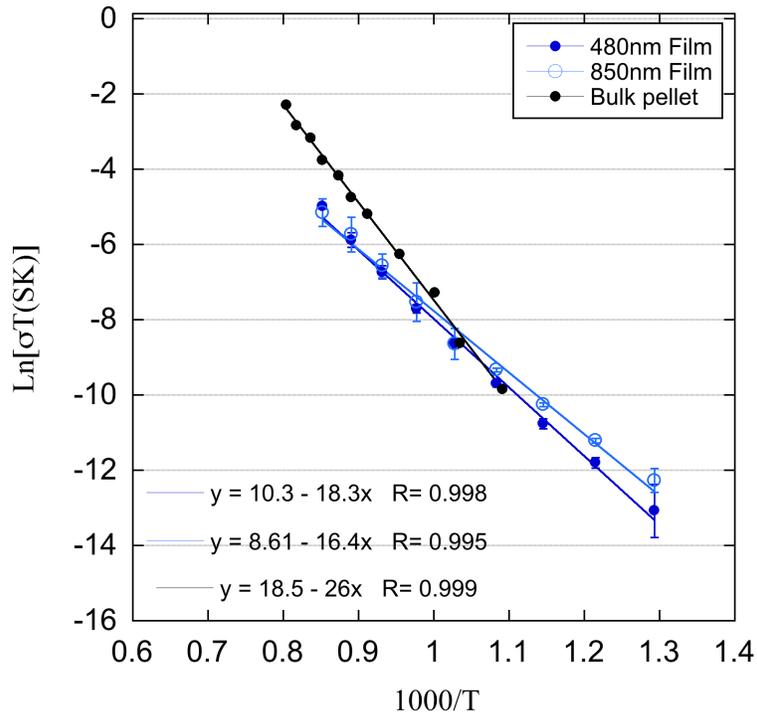

Figure 5: Arrhenius plot of conductivity of bulk polycrystalline GTO and GTO think films of 480 and 850nm thickness.